# Multimedia Satellite Networks and TCP/IP Traffic Transport

Sastri Kota
Lockheed Martin Mission Systems
1260 Crossman Ave, MS:S40
Sunnyvale, CA 94089
e-mail: sastri.kota@lmco.com

Mukul Goyal, Rohit Goyal, Raj Jain
Computer and Information Science Department
The Ohio State University
2015 Neil Ave., Columbus, OH 43210
e-mail: {mukul,goyal,jain}@cis.ohio-state.edu

## ABSTRACT

To meet an increasing demand for multimedia services and electronic connectivity across the world, satellite networks will play an indispensable role in the deployment of global networks. A number of satellite communication systems have been proposed using geosynchronous (GEO) satellites, medium earth orbit (MEO) and low earth orbit (LEO) constellations operating in the Ka-band and above. At these frequencies satellite networks are able to provide broadband services requiring wider bandwidth than the current services at C or Ku-band. Most of the next generation broadband satellite systems will use ATM or "ATM like" switching with onboard processing to provide full two-way services to and from earth stations. The new services gaining momentum include mobile services, private intranets and high data rate internet access carried over integrated satellite-fiber networks. Several performance issues need to be addressed before a transport layer protocol, like TCP can satisfactorily work over satellite ATM for large delay-bandwidth networks. In this paper, we review the proposed satellite systems and discuss challenges such as, traffic management and QoS requirements for broadband satellite ATM networks. The performance results of TCP enhancements for Unspecified Bit Rate over ATM (ATM-UBR+) for large bandwidth-delay environments with various end system policies and drop policies for several buffer sizes are presented.

## 1 INTRODUCTION

The rapid globalization of the telecommunications industry and the exponential growth of the Internet is placing severe demands on global telecommunications. This demand is further increased by the convergence of computing and communications and by the increasing new applications such as Web surfing, desktop and video conferencing. Satisfying this requirement is one of the greatest challenges before telecommunications industry in 21st century. Satellite communication networks can be an integral part of the newly emerging national and global information infrastructures (NII and GII).

### 1.1 Motivation

Satellite communication offers a number of advantages over traditional terrestrial point-to-point networks. These include:
- Wide geographic coverage including interconnection of remote terrestrial networks ("islands")
- Bandwidth on demand, or Demand Assignment Multiple Access (DAMA) capabilities
- An alternative to damaged fiber-optic networks for disaster recovery options
- Multipoint-to-multipoint communications facilitated by the Internet and broadcasting capability of satellites

During the next millennium, wireless satellite systems will play a significant role in meeting telecommunication needs. The next generation satellite systems, often termed "broadband satellite networks" or "multimedia satellite networks," are being developed to provide global, broadband communication services including high data rate Internet access, private Intranets, and TV broadcasting. Some of these systems will offer data communication services at Ka-band and digital broadcasting at Ku-band. Satellite communication networks can interoperate with the current major technology developments, e.g., Internet Protocol (IP) and Asynchronous Transfer Mode (ATM).[1]

### 1.2 Why Ka-band?

Until recently, Ka-band was used for experimental satellite programs in the U.S., Japan, Italy, and Germany. In the U.S, the Advanced Communications Technology Satellite (ACTS) is being used to demonstrate advanced technologies such as onboard processing and scanning spot beams. A number of applications were tested including: distance learning, telemedicine, credit card financial transactions, high data rate computer interconnections, video conferencing and HDTV. The growing congestion of the C and Ku bands and the success of the ACTS program increased the interest of satellite system developers in the Ka-band satellite communications network for exponentially growing Internet access applications. A rapid convergence of technical, regulatory, and business factors has increased the interest of system developers in Ka-band frequencies. Several factors influence the development of multimedia satellite networks at Ka-band frequencies:
- *Adaptive Power Control and Adaptive Coding:* Adaptive power control and adaptive coding technologies have been developed for improved performance, mitigating propagation error impacts on system performance at Ka-band.
- *High Data Rate:* A large bandwidth allocation to geosynchronous fixed satellite services (GSO FSS) and non-geosynchronous fixed satellite services (NGSO FSS) makes high data rate services feasible over Ka-band systems.
- *Advanced Technology:* Development of low noise transistors operating in the 20 GHz band and high power transistors operating in the 30 GHz band have influenced the development of low cost earth terminals. Space qualified higher efficiency traveling-wave tubes (TWTAs) and ASICs development have improved the processing power. Improved satellite bus designs with efficient solar arrays and higher efficiency electric propulsion methods resulted in cost effective launch vehicles.
- *Regulatory Issue:* The orbital congestion at C- and Ku-bands has necessitated the move to Ka-band.
- *Global Connectivity:* Advanced network protocols and interfaces are being developed for seamless connectivity with terrestrial infrastructure.
- *Efficient Routing:* Onboard processing and fast packet or cell switching (e.g., ATM) makes multimedia services possible.
- *Resource Allocation:* Demand Assignment Multiple Access (DAMA) algorithms along with traffic management schemes provide capacity allocation on a demand basis.
- *Small Terminals:* Multimedia systems will use small and high gain antenna on the ground and on the satellites to overcome path loss and gain fades.
- *Broadband Applications:* Ka-band systems, combining traditional satellite strengths of geographic reach and high bandwidth, provide the operators a large subscriber base with scale of economics to develop consumer products.

## 2 PROPOSED MULTIMEDIA SATELLITE SYSTEMS – PARTIAL LIST

In the past three years, interest in Ka-band satellite systems has dramatically increased, with over 450 satellite applications filed with the ITU. In the U.S., there are currently 13 Geostationary Satellite Orbit (GSO) civilian Ka-band systems licensed by the Federal Communications Commission (FCC), compromising a total of 73 satellites. Two Non-Geostationary Orbit (NGSO) Ka-band systems, compromising another 351 satellites, have also been licensed. Eleven additional GSO, four NGSO, and one hybrid system Ka-band application for license and 16 Q/V-band applications have been filed with FCC. Table 1 provides a partial list of proposed satellite systems at Ka-band.[1]

Brief descriptions of these systems are based on FCC filings. However, all these systems are being redesigned to meet their business plans and dynamically changing market demands.

### 2.1 Astrolink

Lockheed Martin's Astrolink system is composed of a space segment and a ground segment. The space segment is made up of an initial constellation of up to five GEO satellites, interconnected by inter-satellite links. This constellation will later be augmented to nine to meet the traffic demand. The ground segment is made of three principal elements: Subscriber terminals located at the customer premises; gateway earth stations that connect the Astrolink system to major customers and Public Switched Network; and Regional Network Control Center that performs subscriber verification, call set-up, and billing.

The Astrolink network architecture is based on the ATM technology to support the integrated voice, data, video, and multimedia services. The system supports 52,000 full duplex circuits per satellite at 64 kbps or 6.6 Gbps per satellite. The user terminal uplinks employ a hybrid multifrequency time division multiple access scheme.

The Astrolink antenna is a multibeam antenna composed of eight reflectors, four transmitters, and four receivers. Each antenna is equipped with a multitude of feed horns capable of multiple spot beams in one or both circular polarizations. Each of the four transmit antennas generates the spot beams at one of the four user uplink frequencies. Each of the four receive antennas generates the congruent receive spot beams.

### 2.2 Spaceway

Hughes has proposed Spaceway System comprising 20 GEO satellites in 15 orbital locations. Spaceway can support 230,000 users worldwide at data rates of 384 kbps. Communication services will be provided at rates of 161 bps to 1.544 Mbps via terminals with antennas in the range of 66 to 200 cm in diameter. Onboard processing and ATM-based switching is used to route the traffic.

### 2.3 GE*Star

GE American Communications, Inc., has proposed a system of nine GEO satellites occupying five orbit locations. GE American proposes to purchase satellites that each produce 44 spot beams for transmitting and receiving, operating in a fourfold frequency reuse pattern. GE*Star plans for a minimum inbound rate of 128 kbps and 24 Mbps information stream (40 Mbps raw data transmission) in the outbound direction. GE*Star strongly considered inter-satellite links.

### 2.4 PanAmSat

PanAmSat was the first private company to offer global services. It now has five operational satellites providing services over the Atlantic, Pacific, and Indian Oceans. Presently, it is capable of providing services to Latin America, Africa, and Central/Eastern Asia. This system does not have inter-satellite links. The first four satellites operate in C- and Ku-band.

## 2.5 Teledesic

Teledesic, originally proposed to consist of 840 LEO satellites, has been redesigned and the number of satellites is reduced to 288. Teledesic supports inter-satellite links. Teledesic has chosen a LEO system based on the argument that GEO propagation delays are a problem for video conferencing and internet access protocols. GEO systems have developed techniques such as "spoofing" to enhance the Internet protocol performance. In addition, the Internet Engineering Task Force (IETF) has developed selective acknowledgments (SACK) and New Reno versions of TCP to improve the performance over GEO satellites. Recently, Teledesic, Motorola, and Boeing announced a teaming agreement. Subsequently, design changes to the Teledesic have not been announced yet.

## 2.6 Celestri

Motorola proposed Celestri merging its previous Millennium and M-Star, employing 63 Ka-band LEO satellites, at 1,400-km altitude, as well as several GEO satellites. The Ka- and V-band payloads will be combined on the 63-satellite Celestri LEO system. The system is designed to offer subscribers very high data rate access from 64 kbps to 155 Mbps. Each satellite will support up to a capacity of 1.83 Gbps. Uplink rates of 2.048, 51.84, and 155.52 Mbps and downlink at 16.384, 51.84, and 155.52 Mbps using demand-assigned frequency division multiplexing/time division multiple access (DA-FDM/TDMA) will be supported.

## 2.7 SkyBridge

SkyBridge was proposed under the leadership of Alcatel. This system is based on a constellations of 80 LEO satellites to deliver global connectivity to over 20 million business and residential users worldwide with performance comparable to that of future terrestrial broadband technologies. The SkyBridge is designed to operate in the Ku-band. The system provides a per user capacity of 20 Mbps on the downlink and up to 2 Mbps on the uplink. Increments in data rates are in 16 Kbps steps, thereby providing the user with "bandwidth on demand." Total worldwide SkyBridge system capacity amounts to over 200 Gbps. Each gateway can handle up to 350,000 users per visible satellite.

# 3 BROADBAND SATELLITE NETWORK ARCHITECTURE: Example

There are several options that drive the broadband satellite network architecture:
- GSO versus NGSO (e.g., LEOs, MEOs)
- No onboard processing or switching
- Onboard processing with ground ATM switching or "ATM like," cell or fast packet switching
- Onboard processing and onboard ATM or "ATM like" fast cell/packet switching

Most of the next generation multimedia satellite systems have in common features like onboard processing, ATM or "ATM-like" fast packet switching, terminals, gateways, common protocol standards, and inter-satellite links. Figure 2 illustrates a broadband satellite network architecture represented by a ground segment, a space segment, and a network control segment. The ground segment consists of terminals and gateways (GWs) which may be further connected to other legacy public and/or private networks. The Network Control Station (NCS) performs various management and resource allocation functions for the satellite media. Inter-satellite crosslinks in the space segment provide seamless global connectivity via the satellite constellation. The network allows the transmission of ATM cells over satellite, multiplexes and demultiplexes ATM cell streams for uplinks, downlinks, and interfaces to interconnect ATM networks as well as legacy LANs.

## 3.1 Gateways (GWs)

The gateways support several protocol standards such as ATM User Network Interface (ATM-UNI), Frame Relay UNI (FR-UNI), Narrow-band Integrated Digital Network (N-ISDN), and Transmission Control Protocol/Internet Protocol (TCP/IP). The gateways interface unit provides external network connectivity. The number and placement of these gateways in both GEO and LEO systems depend on the traffic demand, performance requirements, and other international regulatory issues.

## 3.2 User Terminals (UTs)

The user Terminals Interface Unit (TIU) supports several protocol standards adapting to the satellite network interface. It includes the physical layer function-alities such as channel coding, modulation/demodulation, and other RF functions. Different types of terminals might support transmission rates starting from 16 kbps, 144 kbps, 384 kbps, or even 2.048 Mbps.

## 3.3 Space Segment

The space segment consists of either a GEO or LEO constellation depending on the system design as discussed in Section 3. In any of the multimedia systems, within payloads full onboard processing and ATM or "ATM-like" switching is assumed. The onboard functions include multiplexing/demultiplexing, channel encoding/decoding, packet modulation/demodulation, and formatting. Many of the switching units are under development.

### 3.4 Control Segment

A Network Control Station (NCS) performs various control and management functions, e.g., configuration management, resource allocation, performance manage-ment, and traffic management. The number and location of these NCSs depend on the size of the network, coverage, and other international standards and regulatory issues.

### 3.5 Interfaces

Interconnectivity to the external private or public networks is possible with the support of the standard protocol. For the satellite ATM case, the signaling protocols based on ITU-T Q.2931 can be used when necessary. For other networks, the common channel signaling protocol, e.g., Signaling System No. 7 (SS7), can be used. The other interconnection interfaces between public and private ATM networks are the ATM Inter-Network Interface (AINI), the Public User Network Interface (PUNI) or the Private Network-Network Interface (PNNI), and the default interface between two public ATM networks, namely, the B-ISDN Inter Carrier Interface (B-ICI). However, these interfaces require further modifications to suit the satellite interface unit development. There is a definite need for an integrated satellite-ATM network infrastructure and standards for interfaces and protocols are in development process.

### 3.6 Onboard Processing and Switching

One of the fundamental drivers of the next generation broadband satellite systems is the onboard processing and ATM fast packet/cell switching. Onboard processing involves demodulation and demultiplexing the received signal. The payload performs decoding and encoding, processing the header information, and routing the data, pointing the antennas, buffering, multiplexing, and retransmitting the data on downlink or inter-satellite link. The major reasons for onboard processing include separation of the uplink from the downlink, a gain of approximately 3 dB in performance, and provision of resources on demand. The advantages of onboard processing and switching include:
- Improved error rates by using effective encoding techniques
- Separation of uplink and downlink
- System efficiency can improve from 37% to nearly 99.5% with packet or cell switching
- Delay improvements
- Routing decisions onboard or via intersatellite links
- No end-to-end retransmissions
- Capacity improvements
- Multiple beams with dual polarization

### 3.7 Satellite ATM Technical Challenges

Satellite-ATM systems will play a significant role in achieving global connectivity by interconnecting geographically dispersed ATM networks. These systems will be able to achieve statistical multiplexing gains while maintaining Quality of Service (QoS) requirements. The ATM paradigm is aimed at supporting the diverse requirements of a variety of traffic sources, and providing flexible transport and switching services in an efficient and cost-effective manner. Hence, there is a growing interest in satellite-ATM networks. However, certain design challenges must be addressed before these systems can be deployed. For example, tradeoff analysis needs to be done to implement the traffic management functions in space versus ground segment meeting the weight and power requirements.[2,3,4]

## 4 TCP/IP TRAFFIC TRANSPORT OVER SATELLITE ATM

TCP/IP is the most popular network protocol suite and hence it is important to study how well these protocols perform on long delay satellite links. The main issue affecting the performance of TCP/IP over satellite links is very large feedback delay compared to terrestrial links. The inherent congestion control mechanism of TCP causes source data rate to reduce rapidly to very low levels with even a few packet loss in a window of data. The increase in data rate is controlled by ACKs received by the source. Large feedback delay implies a proportional delay in using the satellite link efficiently again. Consequently, a number of TCP enhancements (NewReno[6,16], SACK[7]) have been proposed that avoid multiple reductions in source data rate when only a few packets are lost. These enhancements also avoid resending packets already received at the destination. It is important to study the effectiveness of these enhancements in achieving better performance on satellite links. The enhancements in end-to-end TCP protocol are called End System Policies. Satellite ATM link performance can also be improved by using intelligent switch policies. The Early Packet Discard policy [11] maintains a threshold $R$ in the switch buffer. When the buffer occupancy exceeds $R$, all new incoming packets are dropped. Partially received packets are accepted if possible. The Selective Drop policy [3] uses per-VC accounting, i.e., keeps track of current buffer utilization of each active UBR VC. A UBR VC is called *active* if it has at least one cell currently buffered in the switch. The total buffer occupancy, $X$, is allowed to grow until it reaches a threshold $R$, maintained as a fraction of the buffer capacity $K$. A fair allocation is calculated for each active VC, and if the VC's buffer occupancy $X_i$ exceeds its fair allocation, its subsequent incoming packet is dropped. Mathematically, in the Selective Drop scheme, an active VC's entire packet is dropped if

$(X > R)$ AND $(X_i > Z \times X/N_a)$

where $N_a$ is the number of active VCs and *Z* is another threshold parameter (*0 < Z <= 1*) used to scale the effective drop threshold.

Also, it is important to know the minimum buffer requirements at the switch in order to avoid cell drop and hence performance loss. Large buffer sizes may exacerbate the problem of large link delay variation associated with multiple-hop satellite links. Since UBR is the cheapest service provided by ATM networks, a majority of Internet Service Providers may use ATM-UBR service for their TCP/IP traffic. In following sections, we discuss the relative impact of end system policies and switch policies on the performance of satellite ATM-UBR+ networks with TCP/IP. The simulation results discussed next were first presented as a contribution[8] to Traffic Management Working Group of ATM Forum in December 1998 and partial results were published in [12].

We performed full factorial simulations[10] using a WWW traffic model (discussed next) with following parameters:

*TCP flavors*: Vanilla (Slow Start + Congestion Avoidance)[5], Fast Retransmit Recovery (Reno)[5], NewReno [6,16] and SACK[7].

*UBR+ drop policies*: Early Packet Drop (EPD)[11] and Selective Drop (SD)[3].

*Propagation delays*: Satellite (Single-hop GEO, multiple-hop LEO/single-hop MEO) and WAN delays.

*Buffer sizes*: We use three buffer sizes approximately corresponding to 0.5, 1, and 2 times the round trip delay-bandwidth products.

### 4.1 WWW Traffic Model

The WWW uses Hypertext Transfer Protocol (HTTP). HTTP uses TCP/IP for communication between WWW client and WWW servers [13]. Modeling of the WWW traffic is difficult because of the changing nature of web traffic. In this section, we outline our model and the inherent assumptions.

#### 4.1.1 Implications of the HTTP/1.1 standard

The main difference between version 1.1 of the Hypertext Transfer Protocol, HTTP/1.1 [9], and earlier versions is the use of persistent TCP connections as the default behavior for all HTTP connections. In other words, a new TCP connection is not set up for each HTTP/1.1 request. The HTTP client and the HTTP server assume that the TCP connection is persistent until a *Close* request is sent in the HTTP Connection header field. Another important difference between HTTP/1.1 and earlier versions is that the HTTP client can make multiple requests without waiting for responses from the server (called *pipelining*). The earlier models were *closed-loop* in the sense that each request needed a response before the next request could be sent.

#### 4.1.2 WWW Server Model

We model our WWW servers as infinitely fast servers getting file requests from WWW clients. The model is an extension of that specified in SPECweb96 benchmark [14]. In our model, a WWW server, on receiving a request from a WWW client sends some data back. The amount of data to be sent (the requested file size) is determined by classifying the client request into one of five classes (Class 0 through Class 4), shown in Table 2. As shown in the table, 20% of the requests are classified as Class 0 requests, i.e., less than 1 KB of data is sent in response. Similarly 28% of the file requests are classified as Class 1 requests and so on. The file size range of each class and the percentage of client requests falling in that class are also shown.

There are nine discrete sizes in each class (e.g. Class 1 has 1 KB, 2 KB, up to 9 KB and Class 2 has 10 KB, 20 KB, through 90 KB and so on.). Within a class, one of these nine file sizes is selected according to a Poisson distribution with mean 5. The model of discrete sizes in each class is based on the SPECweb96 benchmark [14]. There are three key differences from the SPEC model. First, we assume an infinite server, i.e. no processing time taken by server for a client request. Secondly, we created a new class of file sizes (1 MB - 10 MB), which allows us to model file sizes larger than those in the SPEC benchmark. Finally, we had to change the percentage distribution of client requests into server file size classes to accommodate the new class.

The reason for a new class of file sizes is to model the downloading of large software and offline browsing of search results. The percentages of requests falling into each of file size classes have been changed so that average requested file size is around 120 KB, as opposed to 15 KB in SPECweb96 model. We believe the new figure better represents the current WWW traffic scenario. The reason for having 20% of the requests classified as Class 0 requests is explained in next sub-section.

#### 4.1.3 WWW Client Model

The HTTP-model in [15] describes an empirical model of WWW clients based on observations in a LAN environment. Specifically, a typical client is observed to make, on the average, four HTTP GET requests for a single document. Multiple requests are needed to fetch inline images, if any. With the introduction of JAVA scripts in web pages, additional accesses maybe required to fetch the scripts. Therefore, we use five as the average number of HTTP GET requests. In our model, a WWW client makes 1 to 9 requests for a single document, Poisson distributed around a mean of 5. These requests are separated by a random time interval between 100ms to 500 ms. Caching effects at the clients are ignored.

Typically, the first request from an HTTP client accesses the index page (plain text), which is of size 1 KB or less. Since every fifth request is expected to be an index page access, WWW server classifies 20% (= 1/5) of the client requests as Class 0 requests and sends 1 KB or less data in response.

We also model a time lag between batches of requests (presumably for the same document) that corresponds to the time taken by the user to request a new document, as a constant, 10 seconds. While this may be too short a time for a human user to make decisions, it also weights the possibility of offline browsing where the inter-batch time is much shorter.

We do not attempt to model user behavior across different servers. The main purpose of using this simplistic model is to approximate the small loads offered by individual web connections, and to study the effects of aggregation of such small loads on the network.

## 4.2 Simulation Configuration And Experiments

The configuration (Figure 1) consists of 100 WWW clients being served by 100 WWW servers, one server for each client. Both WWW clients and servers use underlying TCP connections for data transfer. The switches implement the UBR+ service with optional drop policies described before. The following subsections describe various configuration, TCP and switch parameters used in the simulations. A client makes 1 to 9 requests (Poisson distributed around a mean of 5) every 10 seconds. Server classifies each client request in one of 5 classes. *Frequency of Access* of a class, indicates how often a client request is classified as belonging to that class. *File Size Range* associated with each class consists of 9 equally spaced discreet sizes. A file size among 9 in the range is chosen according to a Poisson distribution with mean 5. Finally, server sends the file as and when allowed by underlying TCP connection.

### 4.2.1 Configuration Parameters

The configuration consists of 100 WWW client-server connections using TCP for data transfer. Links connecting server/client TCPs to switches have a bandwidth of 155.52 Mbps (149.76 Mbps after SONET overhead), and a one way delay of 5 microseconds. The link connecting the two switches simulates the desired (WAN/LEO/MEO/GEO) link respectively and has a bandwidth of 45Mbps (T3). The corresponding one-way link delays are 5ms (WAN), 100ms (multi hop LEO/single hop MEO) and 275ms (GEO) respectively. Since the propagation delay on the links connecting client/server TCPs to switches is negligible compared to the delay on the inter-switch link, the round trip times (RTTs) due to propagation delay are 10ms, 200ms and 550ms respectively. All simulations run for 100 seconds. Since every client makes a new set of requests every 10 secs, the simulations run for 10 cycles of client requests.

### 4.2.2 TCP Parameters

Underlying TCP connections send data as specified by the client/server applications. A WWW client asks its TCP to send a 128 byte packet as a request to the WWW server TCP. TCP maximum segment size (MSS) is set to 1024 bytes for WAN simulations and 9180 bytes otherwise. TCP timer granularity is set to 100 ms. TCP maximum receiver window size is chosen so that it is always greater than RTT-bandwidth product of the path. Such a value of receiver window ensures that receiver window does not prevent sending TCPs from filling up the network pipe. For WAN links (10 ms RTT due to propagation delay), the default receiver window size of 64K is sufficient. For MEO links (200 ms RTT), RTT-bandwidth product in bytes is 1,125,000 bytes. By using the TCP window scaling option and having a window scale factor of 5, we achieve maximum window size of 2,097,120 bytes. Similarly, for GEO links (550 ms RTT), the RTT-bandwidth product in bytes is 3,093,750 bytes. We use a window scale factor of 6 to achieve maximum window size of 4,194,240 bytes. TCP "Silly Window Syndrome Avoidance" is disabled because in WWW traffic many small segments (due to small request sizes, small file sizes or last segment of a file) have to be sent immediately. It has been proposed in [16] that instead of having a fixed initial SSTHRESH of 64 KB, the RTT-bandwidth product of the path should be used as initial SSTHRESH. In our simulations, we have used the round trip propagation delay - bandwidth product as the initial SSTHRESH value. Hence the initial SSTHRESH values for WAN, MEO and GEO links are 56,250, 1,125,000 and 3,093,750 bytes respectively. The TCP delay ACK timer is NOT set. Segments are ACKed as soon as they are received.

### 4.2.3 Switch Parameters

The drop threshold $R$ is 0.8 for both drop policies – Early Packet Discard (EPD) and Selective Drop (SD). For SD simulations, threshold $Z$ also has a value 0.8. We use three different values of buffer sizes in our experiments. These buffer sizes approximately correspond to 0.5 RTT, 1 RTT and 2 RTT - bandwidth products of the end-to-end TCP connections for each of the three propagation delays respectively. For WAN delays, the largest buffer size is 2300 cells. This is a little more than the 2 RTT - bandwidth product. The reason for selecting 2300 is that this is the smallest buffer size that can hold one complete packet (MSS=1024 bytes) for each of the 100 TCP connections. For WAN, 0.5 RTT and 1 RTT buffers are not sufficient to hold a single packet from all TCPs. This problem will also occur in MEO and GEO TCPs if the number of flows is increased. Some preliminary analysis has shown that the buffer size required for good performance may be related to the number of active TCP connections as well as the RTT-bandwidth product. Further research needs to be performed to provide conclusive results of this effect. Table 3 shows the switch buffer sizes used in the simulations.

### 4.3 Performance Metrics

The performance of TCP is measured by the efficiency and fairness index which are defined as follows. Let $x_i$ be the throughput of the $i$th WWW server ($1 \leq i \leq 100$). Let $C$ be the maximum TCP throughput achievable on the link. Let E be the efficiency of the network. Then, $E$ is defined as

$$E = \frac{\sum_{i=1}^{i=N} x_i}{C}$$

where N = 100 and $\sum x_i$ is sum of all 100 server throughputs.

The server TCP throughput values are measured at the client TCP layers. Throughput is defined as the highest sequence number in bytes received at the client from the server divided by the total simulation time. The results are reported in Mbps.

Due to overheads imposed by TCP, IP, LLC and AAL5 layers, the maximum possible TCP over UBR throughput over a 45Mbps link is much less and depends on the TCP maximum segment size (MSS). For MSS = 1024 bytes (on WAN links), the ATM layer receives 1024 bytes of data + 20 bytes of TCP header + 20 bytes of IP header + 8 bytes of LLC header + 8 bytes of AAL5 trailer. These are padded to produce 23 ATM cells. Thus each TCP segment of 1024 bytes results in 1219 bytes at the ATM layer. Thus, the maximum possible TCP throughput $C$ is 1024/1219 = 84% = 37.80Mbps approximately on a 45Mbps link. Similarly, for MSS = 9180 bytes (on MEO,GEO links), $C$ is 40.39Mbps approximately. Since, the "Silly Window Syndrom Avoidance" is disabled (because of WWW traffic), some of the packets have less than 1 MSS of data. This decreases the value of $C$ a little. However, the resulting decrease in the value of $C$ has an insignificant effect on the overall efficiency metric.

In all simulations, the 45Mbps(T3) link between the two switches is the bottleneck. The maximum possible throughput $C$ on this link is 37.80 Mbps (for WAN) and 40.39 Mbps (for MEO/GEO). The average total load generated by 100 WWW servers is 48 Mbps[1].

We measure fairness by calculating the Fairness Index $F$ defined by:

$$F = \frac{\left(\sum_{i=1}^{i=N} x_i/e_i\right)^2}{N \times \sum_{i=1}^{i=N} (x_i/e_i)^2}$$

where N = 100 and $e_i$ is the expected throughput for connection i. In our simulations, $e_i$ is the max-min fair share that should be allocated to server i. On a link with maximum possible throughput $C$, the fair share of each of the 100 servers is $C/100$. Let $S_i$ be the maximum possible throughput that a server can achieve, calculated as the total data scheduled by the server for the client divided by simulation time.

For all i for which $S_i < C/100$, $e_i = S_i$, i.e., servers that schedule less than their fair share are allocated their scheduled rates. This determines the first iteration of the max-min fairness calculation. These $e_i$'s are subtracted from C, and the remaining capacity is again divided in a max-min manner among the remaining connections. This process is continued until all remaining servers schedule more than the fair share in that iteration, in this case $e_i$ = the fairshare.

### 4.4 Simulation Analysis

In this section, we present a statistical analysis of simulation results for WAN, multiple hop LEO/single hop MEO and GEO links and draw conclusions about optimal choices for TCP flavor, switch buffer sizes and drop policy for these links. The analysis technique we have used here is described in detail in [10]. We give a brief description of these techniques next. The following subsections present simulation results for WAN, LEO/MEO and GEO links respectively.

---

[1] A WWW server gets on average 5 client requests every 10s and sends on average 120 KB of data for each request. This means that on average a WWW server schedules 60KBps i.e. 480Kbps of data. Hence average total load generated by 100 WWW servers is 48Mbps.

**4.4.1 Analysis Technique**

The purpose of analyzing results of a number of experiments is to calculate the individual effects of contributing factors and their interaction. These effects can also help us in drawing meaningful conclusions about the optimum values for different factors. In our case, we have to analyze the effects of the TCP flavor, buffer size and drop policy in determining the efficiency and fairness for WAN, MEO and GEO links. Thus, we have TCP flavor, switch buffer size and drop policy as 3 factors. The values a factor can take are called 'levels' of the factor. For example, EPD and SD are two levels of the factor 'Drop Policy'. Table 4 lists the factors and their levels used in our simulations.

The analysis is done separately for efficiency and fairness, and consists of the calculating the following terms:
- **Overall Mean**: This consists of the calculation of the overall mean 'Y' of the result (efficiency or fairness).
- **Total Variation**: This represents the variation in the result values (efficiency or fairness) around the overall mean 'Y'. *The goal of the analysis to calculate, how much of this variation can be explained by each factor and the interactions between factors.*
- **Main Effects**: These are the individual contributions of a level of a factor to the overall result. A particular main effect is associated with a level of a factor, and indicates how much variation around the overall mean is caused by the level. We calculate the main effects of 4 TCP flavors, 3 buffer sizes, and 2 drop policies.
- **First Order Interactions**: These are the interaction between levels of two factors. In our experiments, there are first order interactions between each TCP flavor and buffer size, between each drop policy and TCP flavor, and between each buffer size and drop policy.
- **Allocation of Variation**: This is used to explain how much each factor contributes to the total variation.
- **Overall Standard Error**: This represents the experimental error associated with each result value. The overall standard error is also used in the calculation of the confidence intervals for each main effect.
- **Confidence Intervals for Main Effects**: The 90% confidence intervals for each main effect are calculated. If a confidence interval contains 0, then the corresponding level of the factor has no significant effect (with 90% confidence) towards the result value (efficiency or fairness). If confidence intervals of two levels overlap, then the effects of both levels are assumed to be similar.

The first step of the analysis is the calculation of the overall mean 'Y' of all the values. The next step is the calculation of the individual contributions of each level 'a' of factor 'A', called the 'Main Effect'. The 'Main Effect' of 'a' is calculated by subtracting the overall mean 'Y' from the mean of all results with 'a' as the value for factor 'A'. The 'Main Effects' are calculated in this way for all the levels of each factor.

We then calculate the interactions between levels of two factors. The interaction between levels of two factors is called 'First-order interaction'. For calculating the interaction between level 'a' of factor 'A' and level 'b' of factor 'B', an estimate is calculated for all results with 'a' and 'b' as values for factors 'A' and 'B'. This estimate is the sum of the overall mean 'Y' and the 'Main Effects' of levels 'a' and 'b'. This estimate is subtracted from the mean of all results with 'a' and 'b' as values for factors 'A' and 'B' to get the 'Interaction' between levels 'a' and 'b'. In our analysis, we calculate only up to 'First-order interactions'. Generally, to get an accurate model of simulation results, 'Main Effects' and 'First-order interactions' are sufficient.

We then perform the calculation of the 'Total Variation' and 'Allocation of Variation'. First, the value of the square of the overall mean 'Y' is multiplied by the total number of results. This value is subtracted from the sum of squares of individual results to get the 'Total Variation' among the results. The next step is the 'Allocation of Total Variation' to individual 'Main Effects' and 'First-order interactions'. To calculate the variation caused by a factor 'A', we take the sum of squares of the main effects of all levels of 'A' and multiply this sum with the number of experiments conducted with each level of 'A'. For example, to calculate the variation caused by TCP flavor, we take the sum of squares of the main effects of all its levels (Vanilla, Reno, NewReno and SACK) and multiply this sum by 6 (with each TCP flavor we conduct 6 different simulations involving 3 buffer sizes and 2 drop policies). In this way, the variation caused by all factors is calculated. To calculate the variation caused by first-order interaction between two factors 'A' and 'B', we take the sum of squares of all the first-order interactions between levels of 'A' and 'B' and multiply this sum with the number of experiments conducted with each combination of levels of 'A' and 'B'.

The next step of the analysis is to calculate the overall standard error for the results. This value requires calculation of individual errors in results and the degrees of freedom for the errors. For each result value, an estimate is calculated by summing up the overall mean 'Y', main effects of the parameter levels for the result and their interactions. This estimate is subtracted from the actual result to get the error '$e_i$' for the result.

If a factor 'A' has '$N_A$' levels, then the total number of degrees of freedom is $\Pi(N_A)$. Thus, for our analysis, the total number of degrees of freedom is $4 \times 2 \times 3 = 24$. The degrees of freedom associated with the overall mean 'Y' is 1. The degrees of freedom associated with 'main effects' of a factor 'A' are '$N_A - 1$'. Thus, degrees of freedom associated with all 'main effects' are $\sum(N_A - 1)$. Similarly, the degrees of freedom associated with first-order interaction between 2 factors 'A' and 'B' are $(N_A - 1) \times (N_B - 1)$. Thus, degrees of freedom associated with all first-order interactions are $\sum(N_A - 1) \times (N_B - 1)$, with the summation extending over all factors. In our analysis, the degrees of freedom associated with all 'main effects' are $3 + 1 + 2 = 6$ and the degrees of freedom associated with all first-order interactions are $(3 \times 1) + (3 \times 2) + (1 \times 2) = 11$.

Since we use the overall mean 'Y', the main effects of individual levels and their first-order interactions to calculate the estimate, the value of the degrees of freedom for errors '$d_e$' is calculated as follows:

$$d_e = \prod(N_A) - 1 - \sum(N_A - 1) - \sum(N_A - 1) \times (N_B - 1)$$

In our case, $d_e = 24 - 1 - 6 - 11 = 6$.

To calculate the overall standard error '$s_e$', the sum of squares of all individual errors '$e_i$' is divided by the number of degrees of freedom for errors '$d_e$' (6 in our case). The square root of the resulting value is the overall standard error.

$$s_e = \sqrt{\left(\sum e_i^2\right) / d_e}$$

Finally, based on the overall standard error, we calculate the 90% confidence intervals for all 'main effects' of each factor. For this purpose, we calculate the standard deviation '$s_A$' associated with each factor 'A' as follows:

$$s_A = s_e \times \sqrt{(N_A - 1) / \prod(N_A)}$$

Here, '$N_A$' is the number of levels for factor 'A' and $\prod(N_A)$ is the total number of degrees of freedom.

The variation around the 'main effect' of all levels of a factor 'A' to get a 90% confidence level is given by the standard deviation '$s_A$' multiplied by $t[0.95, d_e]$, where $t[0.95, d_e]$ values are quantiles of the $t$ distribution. Hence, if '$ME_a$' is the value of the main effect of level 'a' of factor 'A', then the 90% confidence interval for '$ME_a$' is $\{ME_a \pm s_A \times t[0.95, d_e]\}$. The main effect value is significant only if the confidence interval does not include 0.

### 4.5 Simulation Results for WAN links

Table 5 presents the individual efficiency and fairness results for WAN links. Table 6 shows the calculation of 'Total Variation' in WAN results and 'Allocation of Variation' to main effects and first-order interactions. Table 7 shows the 90% confidence intervals for the main effects. A negative value of main effect implies that the corresponding level of the factor decreases the overall efficiency and vice versa. If a confidence interval encloses 0, the corresponding level of the factor is assumed to be not significant in determining performance.

#### 4.5.1 Analysis of Efficiency Values: Results and Observations

The following conclusions can be made from the above tables.

TCP type explains 57.5% of the variation and hence is the major factor in determining efficiency. It can be established from confidence intervals of effects of different TCP types that NewReno and SACK have better efficiency performance than Vanilla and Reno. Since the confidence intervals of effects of SACK and NewReno overlap, we cannot say that one performs better than the other. Confidence intervals for the effects of Vanilla and Reno suggest that Reno performs better than Vanilla.

Buffer size explains 30.24% of the variation and hence is the next major determinant of efficiency. Confidence intervals for effects of different buffer sizes clearly indicate that efficiency increases substantially as buffer size is increased. However, if we look at individual efficiency values, it can be noticed that only Vanilla and Reno get substantial increase in efficiency as buffer size is increased from 1 RTT to 2 RTT.

The interaction between buffer size and TCP type explains 8.99% of the variation. The large interaction is because of the fact that only Vanilla and Reno show substantial gains in efficiency as the buffer size is increased from 1 RTT to 2 RTT. For SACK and NewReno, increasing buffer sizes from 1 RTT to 2 RTT does not bring much increase in efficiency. This indicates that SACK and NewReno can tolerate the level of packet loss caused by a buffer size of 1 RTT.

Though the variation explained by drop policy is negligible, it can be seen that for Vanilla and Reno, SD results in better efficiency than EPD for the same buffer size. This is because for EPD, after crossing the threshold R, all new packets are dropped and buffer occupancy does not increase much beyond R. However for SD, packets of VCs with low buffer occupancy are still accepted. This allows the buffer to be utilized more efficiently and fairly and to better efficiency as well as fairness.

However, for NewReno and SACK, the efficiency values are similar for EPD and SD for same buffer size. This is because NewReno and SACK are much more tolerant of packet loss than Vanilla and Reno. Thus the small decrease in number of packets dropped due to increased buffer utilization does not cause a significant increase in efficiency.

It can be noticed from individual efficiency values that SACK generally performs a little better than NewReno except when buffer size is very low (0.5 RTT). Better performance of NewReno for very low buffer size can be explained as follows. Low buffer size means that a large number of packets are dropped. When in fast retransmit phase, NewReno retransmits a packet for every partial ACK received. However, SACK does not retransmit any packet till *pipe* goes below *CWND* value. A large number of dropped packets mean that not many duplicate or partial ACKs are forthcoming. Hence *pipe* may not reduce sufficiently to allow SACK to retransmit all the lost packets quickly. Thus, SACK's performance may perform worse than NewReno under extreme congestion.

We conclude that SACK and NewReno give best performance in terms of efficiency for WAN links. For NewReno and SACK, a buffer size of 1 RTT seems to be sufficient for getting close to best efficiency with either EPD or SD as the switch drop policy. As discussed before, buffer requirements need to be verified for situations where number of flows is much larger.

### 4.5.2 Analysis of Fairness values: Results and Observations

Buffer size largely determines fairness as 53.55 % of the variation is explained by the buffer size. Confidence intervals for effects of buffer sizes suggest that fairness increases substantially as buffer size is increased from 0.5 RTT to 1 RTT. Since confidence intervals for buffers of 1 RTT and 2 RTTs overlap, it cannot be concluded that 2 RTT buffers result in better performance than 1 RTT buffers.

TCP type is the next major factor in determining fairness as it explains 21.49 % of the variation. Confidence intervals for effects of TCP type on fairness, clearly suggest that NewReno results in best fairness and SACK results in the worst fairness.

SD only increases fairness for low buffer sizes. Overall, both the allocation of variation to drop policy, and confidence intervals for effects of SD and EPD suggest that SD *does not* result in higher fairness when compared to EPD for bursty traffic in WAN links unless buffer sizes are small. This result is interesting and means that *per-flow accounting* to improve fairness will be successful only in presence of sufficiently large buffers.

## 4.6 Simulation Results for MEO links

Table 8 presents the individual efficiency and fairness results for MEO links. Table 9 shows the calculation of 'Total Variation' in MEO results and 'Allocation of Variation' to main effects and first-order interactions. Table 10 shows the 90% confidence intervals for main effects.

### 4.6.1 Analysis of Efficiency values: Results and Observations

TCP flavor explains 56.75% of the variation and hence is the major factor in deciding efficiency value. Non-overlapping confidence intervals for effects of TCP flavors clearly indicate that SACK results in best efficiency followed by NewReno, Reno and Vanilla. However, it should be noticed that difference in performance for different TCP flavors is not very large.

Buffer size explains 21.73% of the variation and hence is the next major determinant of efficiency. Confidence intervals for effects of different buffer sizes indicate that efficiency does increase but only slightly as buffer size is increased. However, Vanilla's efficiency increases by about 5% with increase in buffer size from 0.5 RTT to 2 RTT. The corresponding increase in efficiency for other TCP flavors is around 2% or less. This also explains the large interaction between buffer sizes and TCP flavors (explaining 13.42% of the total variation).

Drop policy does not cause any significant difference in efficiency values.

Thus the simulation results indicate that SACK gives best performance in terms of efficiency for MEO links. However, difference in performance for SACK and other TCP flavors is not substantial. For SACK, NewReno and FRR, the increase in efficiency with increasing buffer size is very small. For MEO links, 0.5 RTT seems to be the optimal buffer size for all non-Vanilla TCP flavors with either EPD or SD as drop policy[2].

### 4.6.2 Analysis of Fairness values: Results and Observations

As we can see from individual fairness values, there is not much difference between fairness values for different TCP types, buffer sizes or drop policies. This claim is also supported by the fact that all 9 main effects have very small values, and for 8 of them, their confidence interval encloses 0. Thus, these simulations do not give us any information regarding fairness performance of different options.

## 4.7 Simulation Results for GEO links

Table 11 presents the individual efficiency and fairness results for GEO links. Table 12 shows the calculation of 'Total Variation' in GEO results and 'Allocation of Variation' to main effects and first-order interactions. Table 13 shows the 90% confidence intervals for main effects.

---

[2] Again this result needs to be verified in presence of much larger number of flows.

### 4.7.1 Analysis of Efficiency values: Results and Observations

TCP flavor explains 69.16% of the variation and hence is the major factor in deciding efficiency value. Confidence intervals for effects of TCP flavors clearly indicate that SACK results in substantially better efficiency than other TCP flavors. Since confidence intervals overlap for NewReno, Reno and Vanilla, one can not be said to be better than other in terms of efficiency.

Buffer size explains 13.65% of the variation and interaction between buffer size and TCP flavors explains 7.54% of the variation. Confidence intervals for 0.5 RTT and 1 RTT buffer overlap, thus indicating similar performance. There is a marginal improvement in performance as buffer size is increased to 2 RTT. Vanilla and Reno show substantial efficiency gains as buffer size is increased from 1 RTT to 2 RTT. There is not much improvement for Vanilla and FRR when buffer is increased from 0.5 RTT to 1 RTT. Hence, in this case, 1 RTT buffer does not sufficiently reduce number of packets dropped to cause an increase in efficiency. However, for a buffer of 2 RTT, the reduction in number of dropped packets is enough to improve Vanilla and Reno's performance.

Drop policy does not have an impact in terms of efficiency as indicated by negligible allocation of variation to drop policy.

From the observations above, it can be concluded that SACK with 0.5 RTT buffer is the optimal choice for GEO links with either of EPD and SD as switch drop policy.

### 4.7.2 Analysis of Fairness values: Results and Observations

The conclusion here is similar to MEO delays. As we can see from individual fairness values, there is not much difference between fairness values for different TCP types, buffer sizes or drop policies. All 9 main effects have very small values, and for 8 of them, their confidence intervals enclose 0. Thus, these simulations do not give us any information regarding relative fairness performance of different options.

## 4.8 Overall Analysis

It is interesting to notice how the relative behavior of different TCP types change as link delay increases. As link delay increases, SACK clearly comes out to be superior than NewReno in terms of efficiency. For WAN delays, SACK and NewReno have similar efficiency values. For MEO delays, SACK performs a little better than NewReno and for GEO delays, SACK clearly outperforms NewReno. The reason for this behavior is that NewReno needs N RTTs to recover from N packet losses in a window whereas SACK can recover much faster and start increasing CWND again. This effect becomes more and more pronounced as RTT increases.

Per-flow accounting scheme SD does not always lead to increase in fairness when compared to EPD. This result can partly be attributed to nature of WWW traffic. SD accepts packets of only under-represented VCs after crossing the threshold R. For sufficient buffer size, many of these VCs are under represented in switch buffer because they do not have a lot of data to send. Thus, SD fails to cause significant increase in fairness. Also, it seems that per-flow accounting is useful only in presence of sufficiently large buffers. It is our intuition that required buffer size for a link is mainly determined by its bandwidth-delay product as well as number of flows. Finding optimal buffer size for given link and traffic conditions remains a research problem.

In summary, our simulation study indicates that as delay increases, the marginal gains of end system policies become more important compared to the marginal gains of drop policies and larger buffers.

## 5 Conclusions

Multimedia satellites networks are the new generation communication satellite systems that will use onboard processing and ATM and/or "ATM-like" switching to provide two-way communications. The proposed satellite or broadband satellite systems operate at Ka-band and above frequencies. Systems using Low Earth Orbit (LEO), Medium Earth Orbit (MEO), and Geosynchronous Earth Orbit (GEO) configurations have been proposed. Several technical challenges need to be surmounted before these systems can be successfully used. In this paper, we have identified such challenges. We have also analyzed design parameters based on end policies and switch parameters for efficient satellite ATM networks. Our analysis indicates that as delay increases, the marginal gains of end system policies become more important compared to the marginal gains of drop policies and larger buffers.

*Table 1. Partial List of Proposed Satellite Systems*

| System | Organization | Altitude (km) | No. of Satellites | System Throughput | On-board Switching/Processing | ISL | Access Scheme |
|---|---|---|---|---|---|---|---|
| Astrolink | Lockheed Martin | 35,786 | 9 | 61 Gb/s | ATM-based | Yes | MF-TDMA |
| Spaceway | Hughes | 35,786 | 20 | 88 Gb/s | ATM-based | Yes | MF-TDMA |
| GE*Star | Ge Americom | 35,786 | 9 | 44 Gb/s | Yes | - | TDMA |
| PanAmsat | Hughes | 35,786 | 9 | Gb/s | Yes | No | TDMA |
| Teledesic | Teledesic Corp. | 1,400 | 288 | 64 Gb/s | Fast Packet SW | Yes | MF-TDMA |
| Celestri** | Motorola | 1,400 | 63 | 116 Gb/s | ATM-based | Yes | FDM/TDMA |
| SkyBridge* | Alcatel | 1,469 | 80 | 200 Gb/s | No | No | MF-TDMA |

*Ku-band;   **Recently merged with Teledesic*

*Table 2 WWW Server File Size Classes*

| Class | File Size Range | Frequency Of Access |
|---|---|---|
| Class 0 | 0 - 1 KB | 20 % |
| Class 1 | 1 KB - 10 KB | 28 % |
| Class 2 | 10 KB – 100 KB | 40 % |
| Class 3 | 100 KB - 1 MB | 11.2 % |
| Class 4 | 1 MB - 10 MB | 0.8 % |

*Table 3. Switch Buffer Sizes Used for Simulations*

| Link Type (RTT) | RTT-bandwidth product (cells) | Switch Buffer Sizes (cells) |
|---|---|---|
| WAN (10ms) | 1062 | 531, 1062, 2300 |
| MEO (200 ms) | 21230 | 10615, 21230, 42460 |
| Single-Hop GEO (550 ms) | 58380 | 29190, 58380, 116760 |

*Table 4. Factors and Levels in Simulations*

| Factor | Level | | | |
|---|---|---|---|---|
| TCP flavor | Vanilla | Reno | NewReno | SACK |
| Switch drop policy | EPD | | SD | |
| Switch buffer size | 0.5 RTT | | 1 RTT | 2 RTT |

*Table 5 Simulation Results for WAN links*

| Drop Policy | TCP Flavor | Buffer = 0.5 RTT | | Buffer = 1 RTT | | Buffer = 2 RTT | |
|---|---|---|---|---|---|---|---|
| | | Efficiency | Fairness | Efficiency | Fairness | Efficiency | Fairness |
| EPD | Vanilla | 0.4245 | 0.5993 | 0.5741 | 0.9171 | 0.7234 | 0.9516 |
| | Reno | 0.6056 | 0.8031 | 0.7337 | 0.9373 | 0.8373 | 0.9666 |
| | NewReno | 0.8488 | 0.8928 | 0.8866 | 0.9323 | 0.8932 | 0.9720 |
| | SACK | 0.8144 | 0.7937 | 0.8948 | 0.8760 | 0.9080 | 0.8238 |
| SD | Vanilla | 0.4719 | 0.6996 | 0.6380 | 0.9296 | 0.8125 | 0.9688 |
| | Reno | 0.6474 | 0.8230 | 0.8043 | 0.9462 | 0.8674 | 0.9698 |
| | NewReno | 0.8101 | 0.9089 | 0.8645 | 0.9181 | 0.8808 | 0.9709 |
| | SACK | 0.7384 | 0.6536 | 0.8951 | 0.8508 | 0.9075 | 0.8989 |

*Table 6 Allocation of Variation for WAN Efficiency and Fairness Values*

| Component | Sum of Squares | | %age of Variation | |
|---|---|---|---|---|
| | Efficiency | Fairness | Efficiency | Fairness |
| **Individual Values** | 14.6897 | 18.6266 | | |
| **Overall Mean** | 14.2331 | 18.3816 | | |
| **Total Variation** | 0.4565 | 0.2450 | 100 | 100 |
| **Main Effects:** | | | | |
| **TCP Flavor** | 0.2625 | 0.0526 | 57.50 | 21.49 |
| **Buffer Size** | 0.1381 | 0.1312 | 30.24 | 53.55 |
| **Drop Policy** | 0.0016 | 0.0002 | 0.34 | 0.09 |
| **First-order Interactions:** | | | | |
| **TCP Flavor-Buffer Size** | 0.0411 | 0.0424 | 8.99 | 17.32 |
| **TCP Flavor-Drop Policy** | 0.0104 | 0.0041 | 2.27 | 1.68 |
| **Buffer Size-Drop Policy** | 0.0015 | 0.0009 | 0.33 | 0.38 |
| **Standard Error, $s_e$ = 0.0156(For Efficiency), 0.0472(For Fairness)** | | | | |

*Table 7 Main Effects and their Confidence Intervals for WAN*

| Factor | Main Effect | | Confidence Interval | |
|---|---|---|---|---|
| | Efficiency | Fairness | Efficiency | Fairness |
| **TCP Flavor:** | | | | |
| Vanilla | -0.1627 | -0.0308 | (-0.1734,-0.1520) | (-0.0632,0.0016) |
| Reno | -0.0208 | 0.0325 | (-0.0315,-0.0101) | (0.0000, 0.0649) |
| NewReno | 0.0939 | 0.0573 | (0.0832,0.1046) | (0.0248, 0.0898) |
| SACK | 0.0896 | -0.0590 | (0.0789,0.1003) | (-0.0914, -0.0265) |
| **Buffer Size:** | | | | |
| 0.5 RTT | -0.1000 | -0.1034 | (-0.1087,-0.0912) | (-0.1299,-0.0769) |
| 1 RTT | 0.0163 | 0.0382 | (0.0076,0.0250) | (0.0117, 0.0647) |
| 2 RTT cells | 0.0837 | 0.0651 | (0.0749,0.0924) | (0.0386, 0.0916) |
| **Drop Policy:** | | | | |
| EPD | -0.0081 | -0.0030 | (-0.0142, -0.0019) | (-0.0217,0.0157) |
| SD | 0.0081 | 0.0030 | (0.0019,0.0142) | (-0.0157, 0.0217) |

*Table 8 Simulation Results for MEO Links*

| Drop Policy | TCP Flavor | Buffer = 0.5 RTT | | Buffer = 1 RTT | | Buffer = 2 RTT | |
|---|---|---|---|---|---|---|---|
| | | Efficiency | Fairness | Efficiency | Fairness | Efficiency | Fairness |
| EPD | Vanilla | 0.8476 | 0.9656 | 0.8788 | 0.9646 | 0.8995 | 0.9594 |
| | Reno | 0.8937 | 0.9659 | 0.9032 | 0.9518 | 0.9091 | 0.9634 |
| | NewReno | 0.9028 | 0.9658 | 0.9105 | 0.9625 | 0.9122 | 0.9616 |
| | SACK | 0.9080 | 0.9517 | 0.9123 | 0.9429 | 0.9165 | 0.9487 |
| SD | Vanilla | 0.8358 | 0.9649 | 0.8719 | 0.9684 | 0.9009 | 0.9615 |
| | Reno | 0.8760 | 0.9688 | 0.8979 | 0.9686 | 0.9020 | 0.9580 |
| | NewReno | 0.8923 | 0.9665 | 0.8923 | 0.9504 | 0.8976 | 0.9560 |
| | SACK | 0.9167 | 0.9552 | 0.9258 | 0.9674 | 0.9373 | 0.9594 |

*Table 9 Allocation of Variation for MEO Efficiency and Fairness Values*

| Component | Sum of Squares | | %age of Variation | |
|---|---|---|---|---|
| | Efficiency | Fairness | Efficiency | Fairness |
| **Individual Values** | 19.3453 | 22.1369 | | |
| **Overall Mean** | 19.3334 | 22.1357 | | |
| **Total Variation** | 0.0119 | 0.0012 | 100 | 100 |
| **Main Effects:** | | | | |
| TCP Flavor | 0.0067 | 0.0003 | 56.75 | 29.20 |
| Buffer Size | 0.0026 | 0.0001 | 21.73 | 7.70 |
| Drop Policy | 0.0001 | 0.0001 | 0.80 | 6.02 |
| **First-order Interactions:** | | | | |
| TCP Flavor-Buffer Size | 0.0016 | 0.0001 | 13.42 | 10.16 |
| TCP Flavor-Drop Policy | 0.0007 | 0.0003 | 6.11 | 22.60 |
| Buffer Size-Drop Policy | 0.0001 | 0.0001 | 0.53 | 6.03 |
| **Standard Error, $s_e$ = 0.0036(For Efficiency), 0.0060(For Fairness)** | | | | |

*Table 10 Main Effects and Their Confidence Intervals for MEO*

| Factor | Mean Effect | | Confidence Interval | |
|---|---|---|---|---|
| | Efficiency | Fairness | Efficiency | Fairness |
| **TCP Flavor:** | | | | |
| **Vanilla** | -0.0251 | 0.0037 | (-0.0276,-0.0226) | (-0.0004,0.0078) |
| **Reno** | -0.0005 | 0.0024 | (-0.0030,0.0019) | (-0.0017,0.0065) |
| **NewReno** | 0.0038 | 0.0001 | (0.0013,0.0062) | (-0.0040,0.0042) |
| **SACK** | 0.0219 | -0.0062 | (0.0194,0.0244) | (-0.0103,-0.0020) |
| **Buffer Size:** | | | | |
| **0.5 RTT** | -0.0134 | 0.0027 | (-0.0154,-0.0114) | (-0.0007,0.0060) |
| **1 RTT** | 0.0016 | -0.0008 | (-0.0005,0.0036) | (-0.0042,0.0026) |
| **2 RTT** | 0.0119 | -0.0019 | (0.0098,0.0139) | (-0.0052,0.0015) |
| **Drop Policy:** | | | | |
| **EPD** | 0.0020 | -0.0017 | (0.0006,0.0034) | (-0.0041,0.0007) |
| **SD** | -0.0020 | 0.0017 | (-0.0034,-0.0006) | (-0.0007,0.0041) |

*Table 11 Simulation Results for GEO Links*

| Drop Policy | TCP Flavor | Buffer = 0.5 RTT | | Buffer = 1 RTT | | Buffer = 2 RTT | |
|---|---|---|---|---|---|---|---|
| | | Efficiency | Fairness | Efficiency | Fairness | Efficiency | Fairness |
| **EPD** | **Vanilla** | 0.7908 | 0.9518 | 0.7924 | 0.9365 | 0.8478 | 0.9496 |
| | **Reno** | 0.8050 | 0.9581 | 0.8172 | 0.9495 | 0.8736 | 0.9305 |
| | **NewReno** | 0.8663 | 0.9613 | 0.8587 | 0.9566 | 0.8455 | 0.9598 |
| | **SACK** | 0.9021 | 0.9192 | 0.9086 | 0.9514 | 0.9210 | 0.9032 |
| **SD** | **Vanilla** | 0.8080 | 0.9593 | 0.8161 | 0.9542 | 0.8685 | 0.9484 |
| | **Reno** | 0.8104 | 0.9671 | 0.7806 | 0.9488 | 0.8626 | 0.9398 |
| | **NewReno** | 0.7902 | 0.9257 | 0.8325 | 0.9477 | 0.8506 | 0.9464 |
| | **SACK** | 0.9177 | 0.9670 | 0.9161 | 0.9411 | 0.9207 | 0.9365 |

*Table 12 Allocation of Variation for GEO Efficiency and Fairness Values*

| Component | Sum of Squares | | %age of Variation | |
|---|---|---|---|---|
| | Efficiency | Fairness | Efficiency | Fairness |
| **Individual Values** | 17.3948 | 21.4938 | | |
| **Overall Mean** | 17.3451 | 21.4884 | | |
| **Total Variation** | 0.0497 | 0.0054 | 100 | 100 |
| **Main Effects:** | | | | |
| **TCP Flavor** | 0.0344 | 0.0008 | 69.16 | 14.47 |
| **Buffer Size** | 0.0068 | 0.0006 | 13.65 | 11.48 |
| **Drop Policy** | 0.0001 | 0.0001 | 0.25 | 2.31 |
| **First-order Interactions:** | | | | |
| **TCP Flavor-Buffer Size** | 0.0037 | 0.0012 | 7.54 | 22.16 |
| **TCP Flavor-Drop Policy** | 0.0025 | 0.0014 | 4.96 | 26.44 |
| **Buffer Size-Drop Policy** | 0.0002 | 0.0001 | 0.41 | 1.45 |
| **Standard Error, $s_e$ = 0.0182(For Efficiency), 0.0139(For Fairness)** | | | | |

*Table 13 Main Effects and Their Confidence Intervals for GEO*

| Factor | Mean Effect | | Confidence Interval | |
|---|---|---|---|---|
| | Efficiency | Fairness | Efficiency | Fairness |
| **TCP Flavor:** | | | | |
| **Vanilla** | -0.0295 | 0.0037 | (-0.0420,-0.0170) | (-0.0058,0.0133) |
| **Reno** | -0.0252 | 0.0027 | (-0.0377,-0.0127) | (-0.0068,0.0123) |
| **NewReno** | -0.0095 | 0.0034 | (-0.0220,0.0030) | (-0.0062,0.0129) |
| **SACK** | 0.0642 | -0.0098 | (0.0517,0.0768) | (-0.0194,-0.0003) |
| **Buffer Size:** | | | | |
| **0.5 RTT** | -0.0138 | 0.0050 | (-0.0240,-0.0036) | (-0.0029,0.0128) |
| **1 RTT** | -0.0099 | 0.0020 | (-0.0201,0.0004) | (-0.0058,0.0098) |
| **2 RTT** | 0.0237 | -0.0070 | (0.0134,0.0339) | (-0.0148,0.0009) |
| **Drop Policy:** | | | | |
| **EPD** | 0.0023 | -0.0023 | (-0.0049,0.0095) | (-0.0078,0.0033) |
| **SD** | -0.0023 | 0.0023 | (-0.0095,0.0049) | (-0.0033,0.0078) |

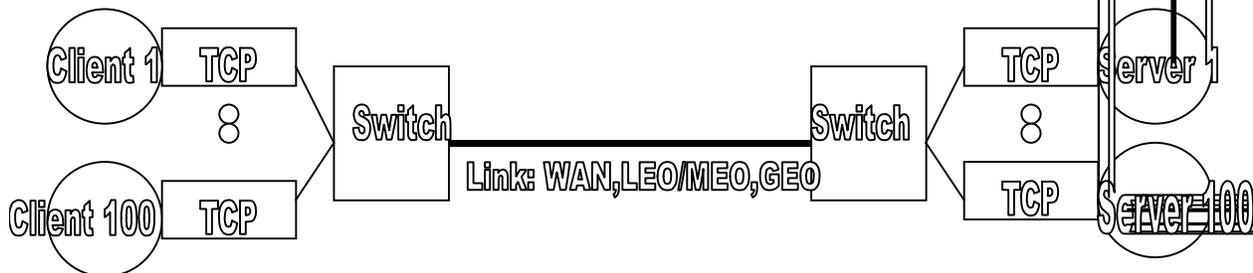

*Figure 1 Simulation Configuration*

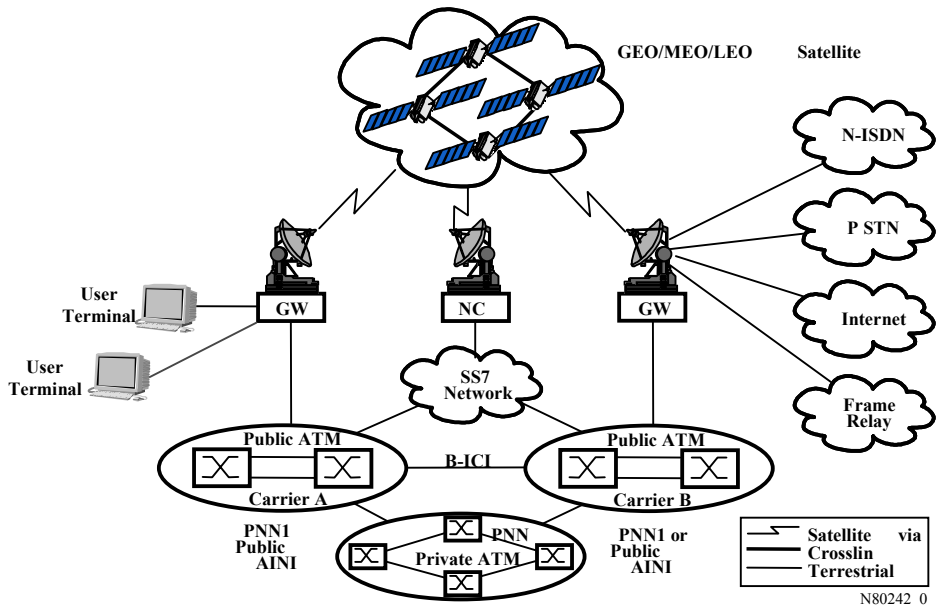

*Figure 2  Broadband Satellite Network Architecture*